\documentclass[12pt]{article} 
\usepackage[utf8]{inputenc}
\usepackage{graphicx} 
\usepackage{float} 
\usepackage[margin=1in,includefoot]{geometry} 
\usepackage{amsmath}
\usepackage{dcolumn}
\usepackage{bm} 
\usepackage{refstyle}
\usepackage{setspace}
\usepackage[numbers,sort&compress]{natbib}
\usepackage{esvect}
\usepackage{subcaption}
\usepackage{physics}
\usepackage{commath}
\usepackage{mathtools}
\usepackage{tabu}
\usepackage[table]{xcolor}
\usepackage{adjustbox}
\usepackage{url}
\usepackage{hyperref}
\usepackage{multirow}
\usepackage{mhchem}
\usepackage{float}
\usepackage{tabularx}
\DeclareUnicodeCharacter{2212}{-}
\begin{document}
\begin{titlepage}
    \begin{flushleft}

\LARGE{\textsc{\textbf{\noindent Simulation of Graphene Nanoplatelets for NO$_2$ and CO Gas Sensing at Room Temperature}}} \\

\vspace{7mm}
\textsc{\large {\textbf{Olasunbo Farinre$^{\dagger}$, Swapnil M. Mhatre$^{\S ,}$$^{\ddagger}$, Albert F. Rigosi$^{\S}$, Prabhakar Misra$^{\dagger*}$}}} \\
    \vspace{5mm}
    \small{$^\dagger  Department$ $of$ $Physics$ $and$ $Astronomy$, $Howard$ $University$, $Washington$, $DC$ $20059$, $USA$} \\
    \vspace{1.5mm}
\small{$^\S Physical$ $Measurement$ $Laboratory$, $National$ $Institute$ $of$ $Standards$ $and$ $Technology$, $Gaithersburg$, $MD$ $20899$, $USA$} \\
    \vspace{1.5mm}
\small{$^\ddagger Graduate$ $Institute$ $of$ $Applied$ $Physics$, $National$ $Taiwan$ $University$, $Taipei$ $10617$, $Taiwan$}

    \vspace{4mm}
    \small{$^{*}$Corresponding Author: Prabhakar Misra; pmisra@howard.edu; Tel. +1 (202) 806 - 6251} 
\end{flushleft}
\end{titlepage}


\newpage
 \begin{abstract}
  \vspace{5mm}
	
	This work reports the modeling and simulation of gas sensors made from pristine graphene nanoplatelets (P-GnPs) using COMSOL Multiphysics software. The mass balance equation was solved while including contributions of electromigration flux. An example GnP-based gas sensor was simulated to undergo exposure to NO$_2$ and CO gases at different concentrations to understand the effects of adsorption. Various electrical properties and the overall sensor responses were also studied as a function of gas concentration in order to determine how viable such sensors could be for target gases. The results herein show that the resistance of the P-GnP-based gas sensor decreases when exposed to NO$_2$ gas whereas an opposite trend is seen when CO gas is used for exposures, ultimately suggesting that the P-GnPs exhibit \textit{p-}type behavior. Sensitivities of 23 \% and 60 \% were achieved when the P-GnP-based gas sensor was exposed to 10 mol/m$^3$ concentration of NO$_2$ and CO at room temperature, respectively. The data heavily suggest that a higher sensitivity towards CO may be observed in future sensors. These simulations will benefit research efforts by providing a method for predicting the behavior of GnP-based gas sensors. 
\end{abstract}

\vspace{5mm}

\newpage

\section{Introduction and Overview}\label{sec:intro}
\
\indent The development of gas sensors with a high sensitivity and selectivity is critical to the identification of toxic gases for emission control, human safety, and environmental preservation. Gas sensors can detect the type, concentration, and constituents of a target gas which in turn produce electrical signals as the corresponding output after a successful detection. Gas sensing materials like two-dimensional (2D) transition metal dichalcogenides (TMDCs) and conducting polymer nanowires such as polyaniline (PANI) have been extensively explored as active layers in gas sensors. These nanomaterials are suitable candidates for gas sensors because they have a high surface area, low cost, and high sensitivity and tunable band gap \cite{wu2013enhanced, song2013conducting, joshi2020two}. However, using TMDCs as gas sensing materials has some drawbacks including poor stability when exposed to water vapor and oxygen \cite{chen20212d} whereas PANI nanowire-based gas sensors typically exhibit loss of conductivity, slow response time, and short life span when in neutral and high pH environments \cite{song2013conducting, pirsa2017chemiresistive}. Nowadays, metal oxides are the most commonly used gas sensing materials because of their high mechanical flexibility, large surface area, good sensitivity, chemical stability, and low production cost \cite{dey2018semiconductor, nurazzi2021frontiers, rzaij2020review, tyagi2016metal}. Despite these advantages, these types of sensors also have some major drawbacks like the PANI nanowire and TMDC-based gas sensors, including poor selectivity, inoperability at room temperature, and short life span \cite{yang2019gas, zaporotskova2016carbon, sahu2020polymer, alzate2020functionalized}. 

\indent The search for lightweight, portable, and highly selective and sensitive gas sensors has led to a significant leap in the fabrication of gas sensors using graphitic nanomaterials such as graphene, carbon nanotubes (CNTs), graphene oxide, and graphene nanoplatelets (GnPs). The specific electrical properties of graphene and its related graphitic counterparts make them attractive for a variety of high-impact applications such as aerospace polymer reinforcement \cite{kalaitzidou2007multifunctional, kim2009multifunctional} electrical metrology \cite{rigosi2019gateless, rigosi2019atypical, rigosi2019quantum}, and most relevantly to this work, gas sensing. These nanomaterials are more attractive for gas sensing when compared to other materials because they have a high thermal stability, making them operable at room temperature \cite{chang2021n}. Also, the possibility of modifying their surface properties by attaching functional groups to their edges and basal plane to enhance the adsorption of target molecules makes them more suitable candidates for gas sensing  \cite{farinre2022comprehensive}. 

\indent Recently, the use of graphene-based materials combined with other materials for gas sensors has been reported. Fei \textit{et al}. \cite{fei2019enhanced} described the fabrication of chemical vapor deposition (CVD)-grown graphene-based gas sensors wherein the surface was decorated with monodisperse polystyrene beads to enhance graphene's ability to sense NO$_2$. Their results show that the sensor's electrical resistance decreases with increasing concentration of NO$_2$, varying from 84.6 $\mu$g/m$^3$ (45 ppb, as defined at standard temperature and pressure) to 188 mg/m$^3$ (100,000 ppb). This decreasing resistance indicates the \textit{p-}type property of both pristine graphene and graphene decorated with polystyrene beads. A decrease in the resistance of the latter was also observed with a sensitivity of 0.8 \% and detection limit of 84.6 $\mu$g/m$^3$ (45 ppb). 

\indent Kostiuk \textit{et al}. \cite{kostiuk2019graphene} reported the gas sensing properties of few-layer graphene nanofilm decorated with palladium nanoparticles, relevant for NO$_2$ and H$_2$ \cite{kostiuk2019graphene}. The film had been synthesized using the modified Langmuir-Schaefer technique and was exposed to increasing concentrations of both gases between room temperature and 200 $^{\circ}$C. Those results demonstrated that the resistance increased when exposed to NO$_2$ and and decreased when exposed to H$_2$. The \textit{p-}type characteristics of graphene were substantiated and agreed with results from Fei \textit{et al}. \cite{fei2019enhanced}. In addition, this groupd reported a response of 26 \% for H$_2$ at 70$^{\circ}$C, whereas a response of about 23 \% was reported at room temperature for a 11.3 mg/m$^3$ (6 ppm) exposure of NO$_2$.

\indent Despite the many interesting properties exhibited by graphitic nanomaterials, there are also some disadvantages they have for gas sensing applications. For instance, CNT-based gas sensors have low sensitivity towards certain gases and a long response time \cite{chang2021n}. Some of the existing problems for graphene include its intrinsically inert nature, its possessed cross sensitivity when exposed to different types of gases \cite{wang2016review}, and its synthesis being too difficult to scale for mass production. Also, graphene oxide has a very high concentration of oxygen functional groups, making it a highly insulating gas sensing material for the fabrication of chemiresistive gas sensors \cite{chang2021n}. GnPs have thus become the alternative gas sensing material because they are easier to synthesize \cite{jimenez2020graphene}, they have potential for large scale production, and they host naturally occurring functional groups, making them chemically reactive unlike graphene. GnPs are nanoscale particles and derivatives of graphite that contain short stacks of graphene per platelet, and the usual size and shape of the platelets make them easy to chemically modify.

\indent One way to reduce long-term resource expenditure, as well as gain a better understanding of the sensing mechanisms of a GnP-based gas sensor, would be to enhance our ability to predict the behavior of the sensor prior to fabrication. For this end, a prototype model of the sensor device was formulated and adapted for use in the COMSOL Multiphysics software \cite{multiphysics1998introduction} (see Acknowledgments). COMSOL was utilized to model and simulate the prototype pristine (P-) GnP-based gas sensor with a high sensitivity, fast response, and quick recovery time. The results presented herein show that P-GnP-based gas sensors exhibit a \textit{p-}type behavior in while sensing a target gas. The sensor was also exposed to increasing concentrations of NO$_{2}$ and CO at room temperature to assess the device's characteristics.

\section{\label{method}Methodology}	  
\subsection{\label{method}Determining Adsorption via Navier-Stokes}	
\
\indent The major processes most relevant to this work and in need of description include a gas sensor's sensitivity to the exposure of NO$_2$ and CO. In the specific case of using P-GnPs as the sensors, adsorption of these two gases can be calculated analytically and modeled with the intent of predicting the change of the electrical resistance exhibited by the sensor. To accomplish these goals, we not only go through the mathematical description, but we also implement COMSOL code to get a numerical prediction of the desired properties. The modeling includes phenomena like the transport of each diluted species by migration in an electric field, the laminar flow of each species, adsorption of those species, and predicted electrical currents through the P-GnPs with newly-modified resistive behaviors. The models rely heavily on partial differential equations (PDE) being solved numerically.

\subsubsection{\label{method}Exposure of P-GnPs to NO$_2$ and CO}	
\
\indent The P-GnP-based gas sensor was exposed to varying concentrations of NO$_2$ and CO by solving the mass balance equation shown below:

\begin{equation} \label{eq:1}
  \frac{\partial c_i}{\partial t}  + \nabla \cdot  \vec {J_i} + \vec u \cdot \nabla c_i = R_i \\
\end{equation} 
\begin{equation} \label{eq:2}
\vec {J_i} = -D_{i} \nabla c_{i} - Z_{i} U_{m,i} F c_{i} \nabla{V}
\end{equation} 

\noindent Here, c$_i$ is the concentration of NO$_2$ and CO gas species, $\vec {J_i}$ is the diffusive flux vector (eq. (2)),  $\vec {u}$ is the fluid velocity, which will be computed from the Navier Stokes equation described later, $R_i$ is the molar flux, $D_i$ is the diffusion coefficient of NO$_2$ and CO gas species, $Z_i$ is the charge number of NO$_2$ and CO gas species, $U_{m,i}$ is the ionic mobility (Nernst-Einstein relation in eq. (3)), $F$ is Faraday's constant, and $V$ is the electric potential.

\begin{equation} \label{eq:3}
 U_{m,i} =  \frac{-D_i}{RT}  \\
 \end{equation} 
 
\noindent Here, $R$ is the molar gas constant (J/(K$\cdot$mol)) and $T$ is the temperature (K). \\

\indent The Navier-Stokes equation is shown below (eq. (4)), and it governs the motion of fluids. It can further be regarded as Newton's second law of motion for fluids and was used to describe the flow of gases in a model cylindrical chamber in which the P-GnP-based sensor resides.

\begin{equation} \label{eq:4}
\underbrace{\rho (\frac{\partial \vec u}{\partial t} + \vec u \cdot \nabla \vec u)}_\text{term 1}  = \underbrace{-\nabla{P}}_\text{term 2} + \underbrace{\nabla \cdot (\mu (\nabla \vec u + (\nabla \vec u)^{T}) -\frac{2}{3} \mu (\nabla \vec u)\vec I}_\text{term 3} + \underbrace{\vec F}_\text{term 4}
\end{equation} 

\noindent Here, $P$,  $\rho$, $\mu$ and $\vec I$ are the fluid pressure, fluid density, and fluid dynamic viscosity, and identity tensor, respectively. The first term corresponds to the inertial forces, whereas the second, third, and fourth term correspond to the pressure forces, viscous forces, and additional external forces applied to the fluid. In the case of an incompressible Newtonian fluid, eq. (4) may be simplified to yield eq. (5) because $\nabla \cdot \vec u$ = 0.

\begin{equation} \label{eq:5}
\underbrace{\rho (\frac{\partial \vec u}{\partial t} + \vec u \cdot \nabla \vec u)}_\text{term 1}  = \underbrace{-\nabla{P}}_\text{term 2} + \underbrace{\mu \nabla^2 \vec u}_\text{term 3} + \underbrace{\vec F}_\text{term 4}
\end{equation} 

\noindent These equations are always solved together with the continuity equation shown in equation (6), which is an analog for mass conservation.

\begin{equation} \label{eq:6}
\frac{\partial \rho}{\partial t} + \nabla \cdot (\rho \vec u) = 0
\end{equation} 

\subsubsection{\label{method}Adsorption of NO$_2$ and CO}	
\
\indent The adsorption and desorption of NO$_2$ and CO at the surface is described by:

\begin{gather}
\mathrm{NO_2 (gas)} \xrightleftharpoons[k_r]{k_f} \mathrm{NO_2 (surface)}  \\
\mathrm{CO (gas)} \xrightleftharpoons[k_r]{k_f} \mathrm{CO (surface)}
\end{gather} 

\noindent Where $k_f$ and  $k_r$ are forward and reverse rate constants and all molecules shown in the reaction are in the gaseous phase and adsorbed by the active layer. The rate of adsorption of NO$_2$ molecules on the active layer is proportional to both the fraction of free surface sites ($1-\theta$) and the rate at which target molecules strike the surface of the active layer. The rate of desorption of the target gases from the active layer is shown in equation (9).

\begin{gather} 
r_{ads} = k_{ads} c(\Gamma_s - c_s) \\
r_{des} = k_{des} c_s
\end{gather} 

\noindent Where $k_{ads}$ = $k_f$RT/$\Gamma_s$ is the rate constant for the adsorption reaction (m$^3$/(mol$\cdot$S)) and $k_{des}$ = $k_r$/$\Gamma_s$ is the rate constant for the desorption reaction (1/S), $\Gamma_s$ is the active site concentration (mol/m$^2$), $c_s$ is the surface concentration and $c$ is the concentration of NO$_2$ molecules in gas phase.

\indent The equation including the surface diffusion and the reaction rate expression for the formation of the adsorbed NO$_2$ molecules on the surface, $c_s$ is:

\begin{equation} \label{eq:10}
\frac{\partial c_s}{\partial t} + \nabla \cdot (-D_s\nabla c_s) = r_{ads} - r_{des}
\end{equation} 

\noindent Where $D_s$ is the surface diffusivity of the active layer. Substituting equations (8) and (9) into equation (10) gives the equation for the transport and reaction on the surface:

\begin{equation} \label{eq:10}
\frac{\partial c_s}{\partial t} + \nabla \cdot (-D_s\nabla c_s) = k_{ads} c(\Gamma_s - c_s) - k_{des} c_s
\end{equation} 

\indent The initial conditions set the concentration of the adsorbed NO$_2$ molecules to be zero at the beginning of the process, $c_s$ = 0 and the concentration in the bulk at $t = 0$ to be equal to the concentration $c$, $c = c_0$.

\subsubsection{\label{method}Electrical Properties of the Active Layer}	
\
\indent When the P-GnP-based sensor has its active layer exposed to NO$_2$ and CO molecules, reactions between the adsorbed molecules and P-GnPs take place, resulting in changes of the sensor's resistance. The continuity equation shown in eq. (13) which expresses local charge conservation is used to describe effects of the interactions between the target gases and P-GnPs on the resistance of the sensor.
\begin{gather}
\nabla \cdot \vec J = -\frac{\partial \rho}{\partial t} \\
\vec J = \sigma \vec E + \vec J_e  \\
\vec E = -\nabla V
\end{gather} 

\noindent Where $\vec J$, $\rho$, $\sigma$, $\vec E$, and $\vec J_e$ are the current density, electric charge density, electrical conductivity, electric field intensity and externally generated current, respectively.

\subsubsection{\label{method}Boundary Conditions}	
\
\indent The boundary conditions of the sensor applied at different domains for solving the mass balance and adsorption on active layer are given below:
\begin{enumerate}
	\item Boundary Conditions for Entry and Exit of NO$_2$ and CO molecules: A ``no slip" boundary condition is applied on the inner-wall side of the gas chamber where the velocity $\vec u$ is zero. Also, the boundary conditions due to the normal inflow velocity of NO$_2$ and CO molecules at the inlet (U$_{inlet}$) and pressure of the outlet (P$_{outlet}$) are:
\begin{gather}
U_{inlet} = 0.05 \hspace{1mm} m/s  \\
P_{outlet} = 0 \hspace{1mm} Pa  
\end{gather} 	
	
	\item Boundary Conditions for Adsorption: The mass balance of the NO$_2$ and CO molecules in the bulk and on the surface are coupled and is obtained as a boundary condition in the bulk's mass balance. This condition sets the flux of concentration of NO$_2$ and CO, $c$, at the boundary equal to the rate of the surface reaction which gives a convection-diffusion equation.
	
	\begin{equation} \label{eq:10}
\frac{\partial c}{\partial t} + \nabla \cdot (-D\nabla c + c\vec u) = 0
\end{equation} 

\noindent where $D$ is the diffusivity of the reacting species. \\
\indent Also, the boundary conditions of the adsorbed NO$_2$ and CO molecules on the active layer are insulating conditions according to eq. (19):

\begin{equation} \label{eq:10}
\vec n \cdot  (-D_s\nabla c_s) = 0
\end{equation} 

\noindent Where $D_s$ is the surface diffusivity. For the bulk NO$_2$ and CO molecules, the boundary condition at the active layer is coupled with the rate of reaction at the active layer, the flux of the reacting species and concentration of adsorbed and bulk NO$_2$ and CO molecules.

\begin{equation} \label{eq:10}
\vec n \cdot  (-D\nabla c + c\vec u) = -k_{ads} c(\Gamma_s - c_s) + k_{des} c_s
\end{equation} 
\noindent The other boundary conditions for the bulk NO$_2$ and CO molecules problem are:
\begin{gather}
inlet: c = c_0 \\
outlet: \vec n \cdot  (D\nabla c) = 0 \\
insulation: \vec n \cdot  (-D\nabla c + c\vec u) = 0
\end{gather} 

\end{enumerate}

\subsection{\label{method}Structure of the Simulated P-GnP-Based Gas Sensor}	
\
\indent The P-GnP-based gas sensor consists of three parts: the substrates, the electrodes and the active layer.

\begin{enumerate}
	\item The substrates: Silicon and silicon dioxide (Si/SiO$_2$) materials were used as substrates for modeling the sensor becasue they are the widely used substrates for the fabrication of graphene-based devices. The detailed parameters of the substrates required for the simulation are shown in table 1.
	\item The electrodes: The material used for the electrodes in the model is gold with its properties highlighted in table 2.
	\item The active layer: P-GnPs is the active layer in the sensor upon which the adsorption of NO$_2$ and CO molecules takes place and its properties are shown in table 3.
	\end{enumerate}

\begin{figure}[H]
\centering
\captionsetup{justification=centering}
\includegraphics[width=7in]{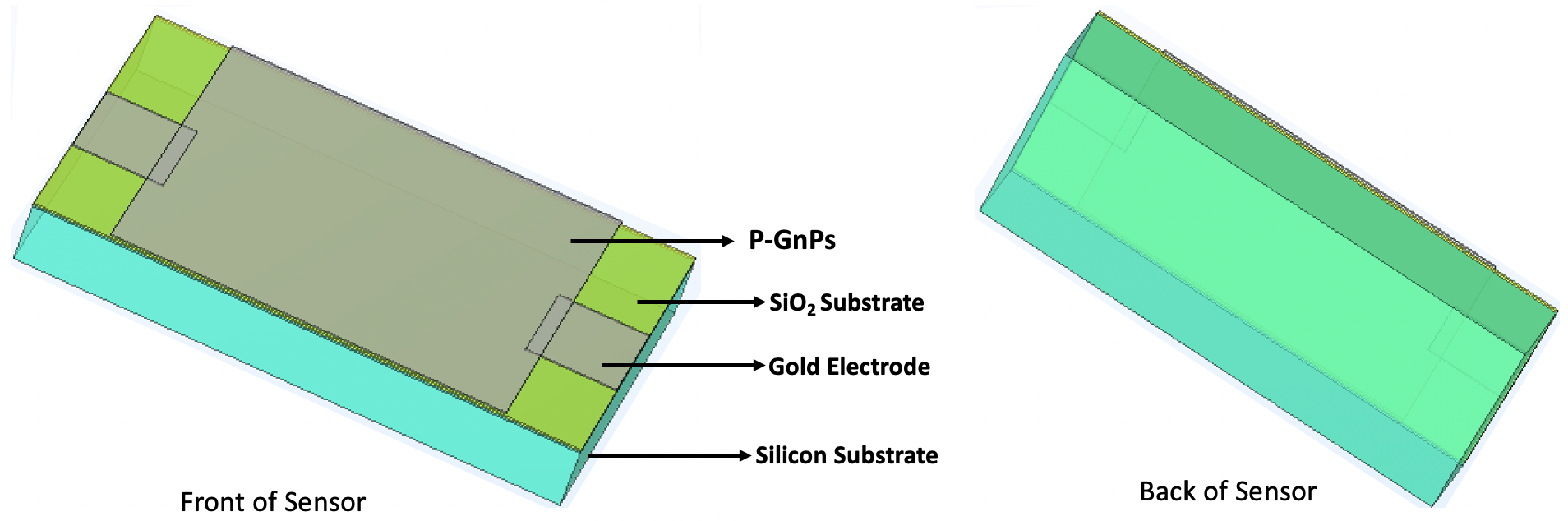}
\caption{\label{figure_1} Structure of P-GnPs-based Gas Sensor}.
\end{figure}

\begin{table}[H]
\centering
\captionsetup{justification=centering}
\caption{\label{t1_1} Properties of Si/SiO$_2$ Substrates}
\begin{tabular}{p{1.5cm}|p{1cm}|p{1cm}|p{1.3cm}|p{2.5cm}|p{2.2cm}}
\hline
\hline
Material & Width (m) & Length (m) & Thickness (m) & Electrical Conductivity (S/m) & Relative permittivity\\
\hline
Silicon & 9x10$^{-3}$ & 6x10$^{-3}$ & 1x10$^{-3}$ & 1x10$^{3}$ \cite{reference.wolfram_2021_elementdata} & 11.7 \cite{honsberg2019photovoltaics}\\  
\hline
Silicon dioxide (SiO$_2$) & 9x10$^{-3}$ & 6x10$^{-3}$ & 3x10$^{-5}$ & 1x10$^{-10}$ \cite{gauthier1995engineering} & 3.9 \cite{muller1986device} \\
\hline
\hline
\end{tabular}
\end{table}

\begin{table}[H]
\centering
\captionsetup{justification=centering}
\caption{\label{t1_1} Properties of Gold Electrodes}
\begin{tabular}{p{1.5cm}|p{1cm}|p{1cm}|p{1.3cm}|p{2.5cm}|p{2.2cm}}
\hline
\hline
Material & Width (m) & Length (m) & Thickness (m) & Electrical Conductivity (S/m) & Relative permittivity\\
\hline
Gold & 1.5x10$^{-3}$ & 1.7x10$^{-3}$ & 2x10$^{-5}$ & 5x10$^{7}$ \cite{seitz1937modern} & 6.9 \cite{shklyarevskii1973separation}\\  
\hline
\hline
\end{tabular}
\end{table}

\begin{table}[H]
\centering
\captionsetup{justification=centering}
\caption{\label{t1_1} Properties of Active Layer}
\begin{tabular}{p{1.5cm}|p{1cm}|p{1cm}|p{1.3cm}|p{2.5cm}|p{2.2cm}}
\hline
\hline
Material & Width (m) & Length (m) & Thickness (m) & Electrical Conductivity (S/m) & Relative permittivity\\
\hline
P-GnPs & 6.5x10$^{-3}$ & 6.0x10$^{-3}$ & 2x10$^{-5}$ & 766.87  & 37.9 \cite{wang2014dielectric}\\  
\hline
\hline
\end{tabular}
\end{table}

\subsection{\label{method}Electrical Measurements of P-GnPs}	
\
\indent Four trials of Current-Voltage ($I$-$V$) measurements were taken using the Alessi REL-4100A analytical probe station (see Acknowledgments) with 1 $\mu$m resolution in all directions. Electrical contacts were fabricated onto compressed pucks composed of P-GnPs to allow for the application of electrical current. 

\indent The $I$-$V$ characteristics of the P-GnP samples were determined within the range of -0.1 V to 0.1 V at room temperature. These results are shown in Figure 1. The average resistance of P-GnPs was found to be 0.2 $\Omega$ and the resistance was used to calculate its conductivity which was found to be 766.87 S/m.

\begin{figure}[H]
\centering
\captionsetup{justification=centering}
\includegraphics[width=5in]{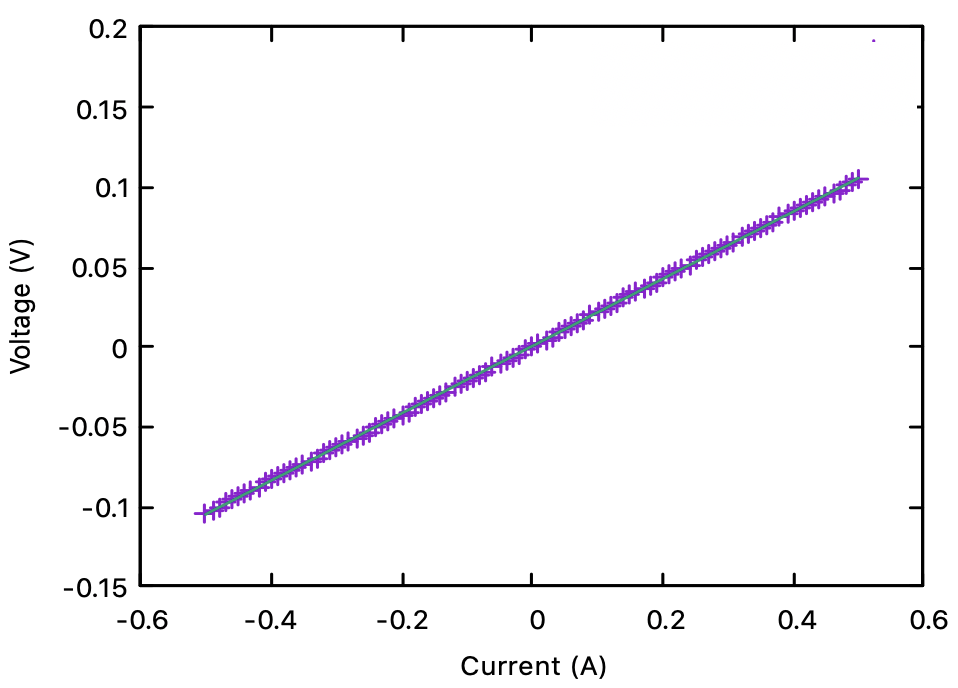}
\caption{\label{figure_1} Current-voltage characteristics of P-GnPs.}
\end{figure}

\section{\label{method}Simulation Results and Discussion}	  
\subsection{\label{method}Exposure of P-GnP-Based Sensors to NO$_2$ and CO}	
\
\indent A cylindrical gas chamber of diameter and height, $d=h=0.02$ m shown in Figure 3 was designed to expose the sensor to NO$_2$ and CO gases. The gas chamber consists of two outlets of radius $r=0.003$ m each located at the bottom of the chamber and one inlet of radius $r=0.005$ m located at the top of the chamber. The inlet serves as an entry point for the gases whereas the two outlets serve as an exit point for the gases. The sensor was placed in the middle of the chamber to allow for uniform and easy adsorption of the molecules on its surface. The base resistance of the P-GnP-based sensor is determined by exposing it to fresh air through the inlet after which air is desorbed from the sensor and replaced with NO$_2$ or CO gas as shown in Figure 3. The base resistance of the P-GnP-based sensor when exposed to air was found to be about 2.13 x 10$^{3}  \hspace{1mm}\Omega$, which agrees well with the experimental resistance of a graphene-based sensor on Si/SiO$_2$ substrates \cite{davaji2017patterned}.

\begin{figure}[H]
\centering
\includegraphics[width=6.8in, height=3.4in]{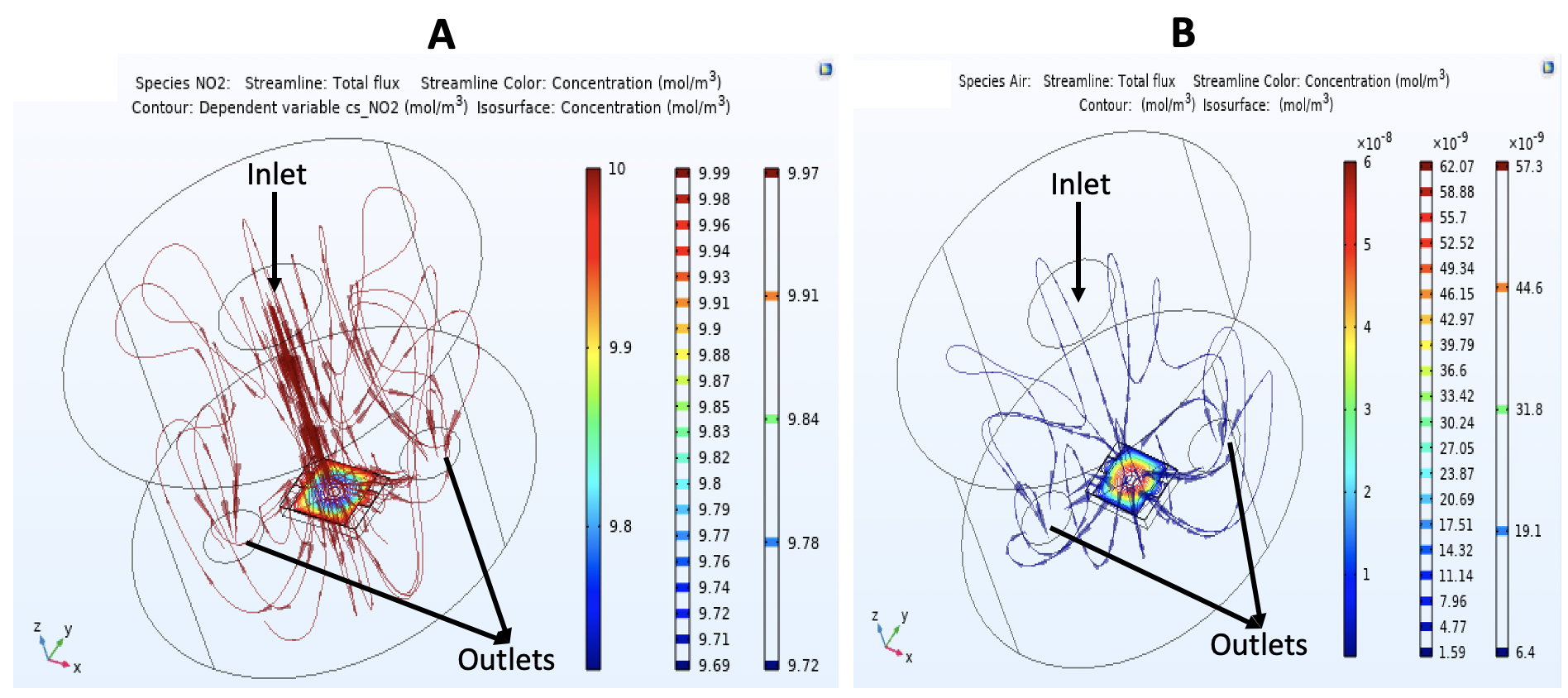}
\caption{\label{figure_1}(A) Adsorption of NO$_2$, and (B) desorption of air from the surface of P-GnP-based gas sensor.}
\end{figure}

\subsection{\label{method}Resistance of Exposed P-GnP-Based Sensors}	
\
\indent Figure 4  shows the resistance of the sensor exposed to NO$_2$ and CO concentrations of 10 mol/m$^3$ to 70 mol/m$^3$  at room temperature. A rapid decrease and increase in the resistance is observed when NO$_2$ and CO gases, respectively, are introduced into the gas chamber for adsorption. When NO$_2$, an oxidizing gas, is adsorbed on the surface of the P-GnPs, electrons are transferred from the P-GnPs to NO$_2$ as shown in eq. (24). This transfer enhances the concentration of holes, thereby decreasing the resistance of the sensor. On the other hand, adsorption of CO, a reducing gas, introduces electrons to the P-GnPs as shown in the chemical equations (25) and (26). The effect becomes opposite from the NO$_2$, and the concentration of holes becomes depleted, increasing the resistance of the sensor. These overall behaviors demonstrate the \textit{p-}type behaviour of P-GnPs \cite{tang2021review}. The initial resistance of the sensor is re-established when fresh air is introduced into the gas chamber and the sensor's resistance reaches equilibrium.
\begin{gather} \label{eq:10}
NO_{2} (gas) + e^-  \longrightarrow  NO_{2}^- (ads)   \\
CO (gas) + O^- (ads)  \longrightarrow  CO_{2} (gas) +  e^-   \\
CO  + 2O^-   \longrightarrow  CO_{3}^{2-}  \longrightarrow  CO_{2} + \frac{1}{2}O_2 + 2e^-
\end{gather}

\begin{figure}[H]
\centering
\includegraphics[width=6.8in, height=3.5in]{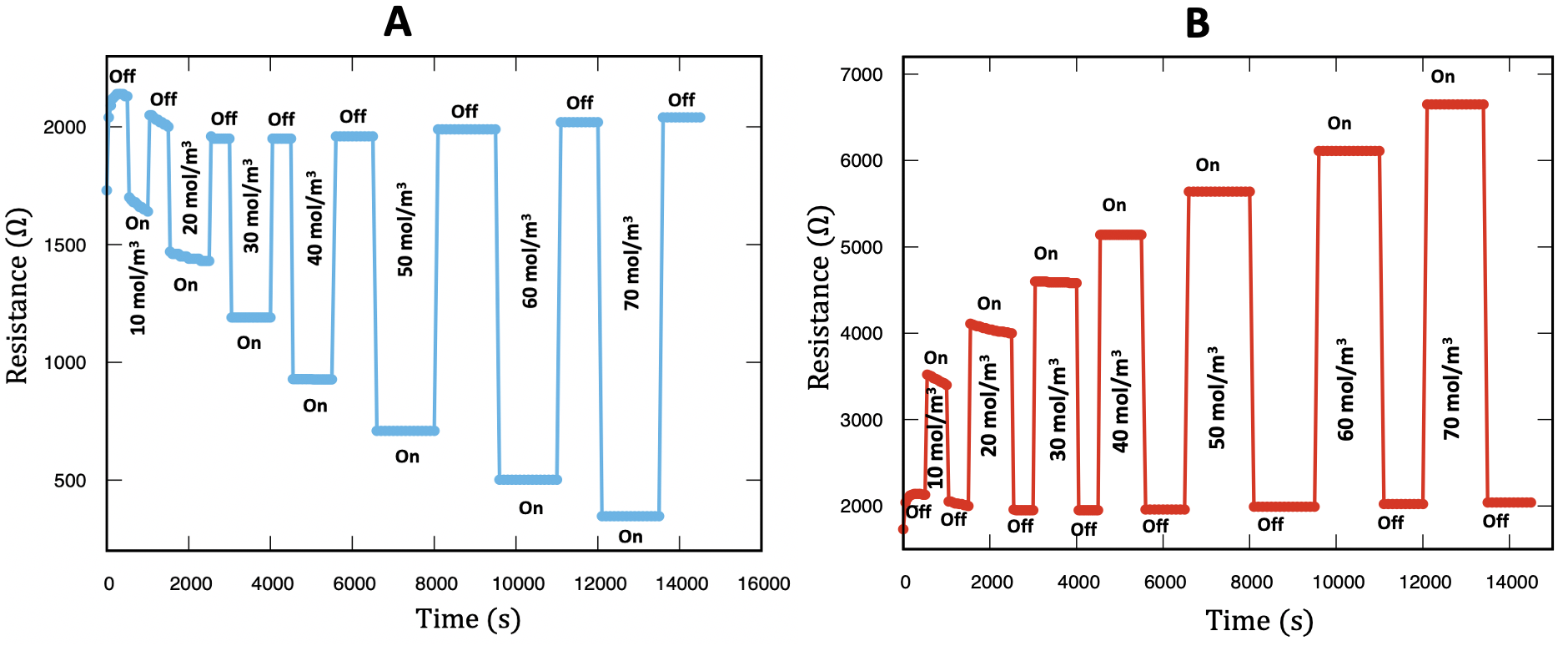}
\caption{\label{figure_1} The time-resistance relation of the simulated P-GnP-based sensor at room temperature when exposed to (A) NO$_{2}$ and (B) CO.}
\end{figure}

\subsection{\label{method}Sensitivity of Exposed P-GnP-Based Sensors}	
\
\indent The sensitivity of the P-GnP-based sensor was calculated using the formula: 

\begin{equation} \label{eq:10}
\frac{R_g - R_a}{R_a} \times 100
\end{equation} 

 \noindent Where $R_a$ is the resistance of the sensor in air and $R_g$ is the resistance of the sensor in NO$_{2}$ or CO gas. The response and recovery cutoff times are defined by a threshold value 90 \% of its equilibrium value after the target has been injected and removed from the gas chamber, respectively \cite{utari2020wearable}. The response of a sensor generally has negative or positive values when exposed to oxidizing or reducing gases depending on the type of conductivity of the sensing medium \cite{kostiuk2019graphene}. Figure 5 illustrates the time-dependent sensitivity curves for the P-GnP-based sensor exposed to different concentrations ranging from 10 mol/m$^3$ to 40 mol/m$^3$ of NO$_{2}$ and CO. The sensor was predicted to achieve fast response and recovery times of about 10 s when exposed to all concentrations of NO$_{2}$ and CO.
 
\begin{figure}[H]
\centering
\includegraphics[width=6.8in, height=3.4in]{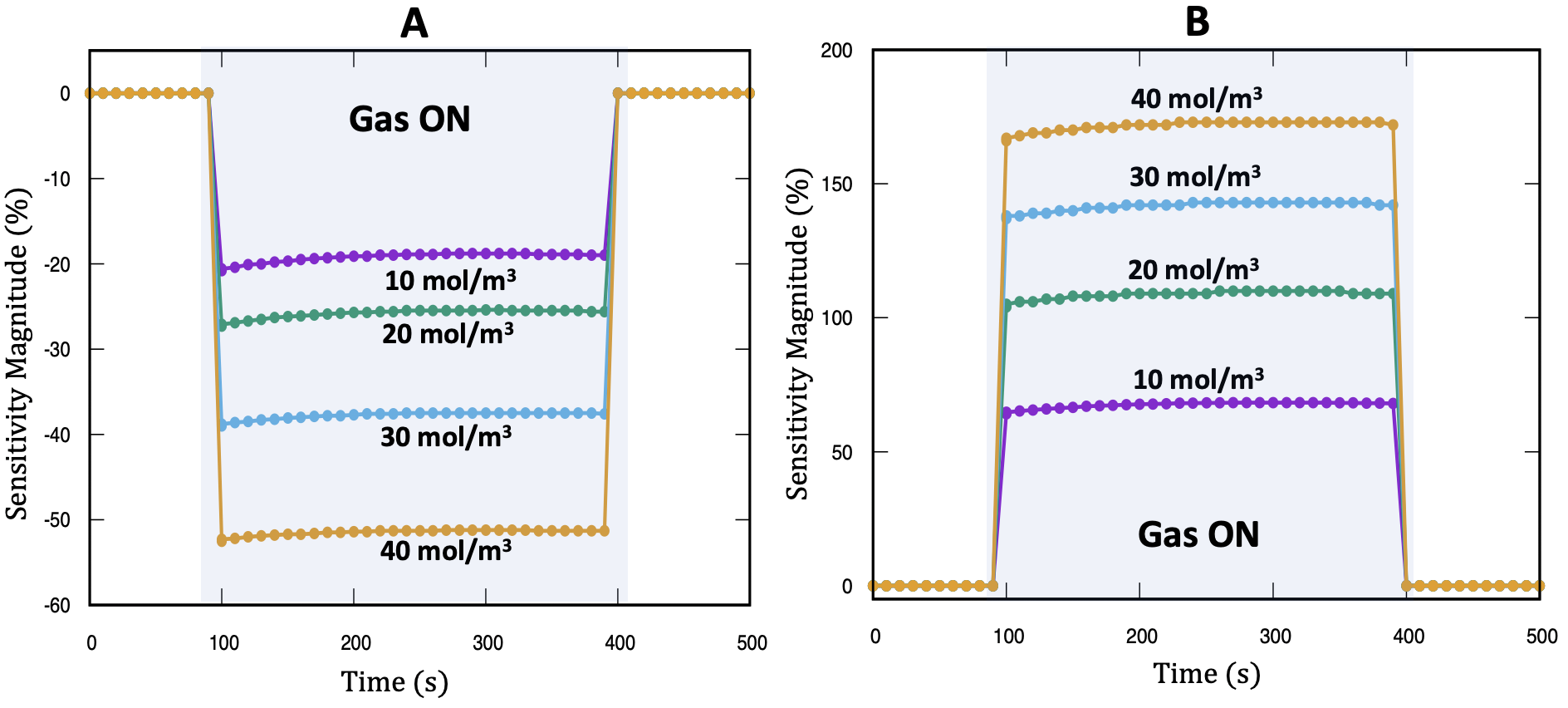}
\caption{\label{figure_1} Response and recovery of the sensor as a function of time and concentration for (A) NO$_{2}$ and (B) CO at room temperature.}
\end{figure}

\
\indent  The sensitivity noticeably increased with an increase in the gas concentration (see Figure 6). When the concentration of NO$_{2}$ increased from 10 mol/m$^3$ to 70 mol/m$^3$, the sensitivity gradually increased from 23 \% to 84 \%, whereas the sensitivity for the CO case increased from 60 \% to 212 \%, indicating that this gas is more reactive with the sensor than the former.

\begin{figure}[H]
\centering
\includegraphics[width=5in, height=4in]{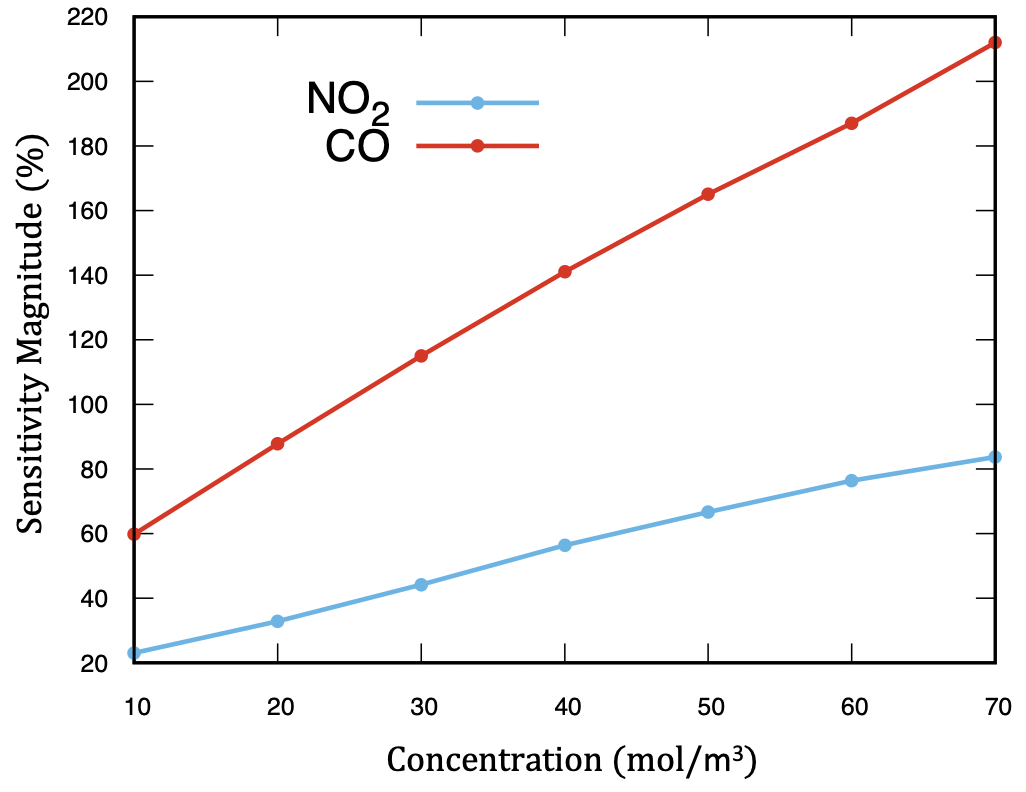}
\caption{\label{figure_1} Sensitivity of simulated P-GnP-based sensors exposed to NO$_{2}$ and CO at room temperature.}
\end{figure}

\section{Conclusions}\label{sec:intro}
\
\indent Future improvements to modeling could include the use of Kwant, notable for its usefulness in quantum transport of related 2D materials [36]. In summary, a gas sensor based on P-GnPs for the detection of NO$_{2}$ (oxidizing) and CO (reducing) gases was modeled and simulated using COMSOL Multiphysics software \cite{multiphysics1998introduction}. These simulation results reveal that the resistance behaviors of the P-GnP-based sensor before and after exposure to the target gases is consistent with observations made in similar devices, like a fabricated graphene-based sensor on Si/SiO$_2$. Additionally, the P-GnP-based sensor is predicted to have a much higher sensitivity towards CO when compared to NO$_{2}$ and may show approximate linearity in its sensitivity as a function of concentration. These sensors were predicted to have fast response and recovery times towards both CO and NO$_{2}$ target gases. Furthermore, the modeling techniques implemented by this study may be later developed to include different types of active layers (like functionalized layers) and varying thicknesses. The natural expansion of this work to include functionalized P-GnP and the inclusion of experimental device fabrication and testing are both warranted to verify the accuracy of these models.

\section*{Author Contributions and Acknowledgments}\label{sec:intro}
\
\noindent  S. M. M. supported with performing the electrical measurements, A. F. R. and P. M. assisted with the analyses, support, and general project oversight. 

O.F. and P.M. would like to acknowledge the financial support from the National Science Foundation Excellence in Research (NSF Award No. DMR 2101121).

Commercial equipment, instruments, and materials are identified in this paper in order to specify the experimental procedure adequately. Such identification is not intended to imply recommendation or endorsement by the National Institute of Standards and Technology or the United States government, nor is it intended to imply that the materials or equipment identified are necessarily the best available for the purpose. The authors declare no competing interests.

\clearpage
 
\section*{References}

[1]	Zuquan Wu, Xiangdong Chen, Shibu Zhu, Zuowan Zhou, Yao Yao, Wei Quan, and Bin Liu. Enhanced sensitivity of ammonia sensor using graphene/polyaniline nanocomposite. Sensors and Actuators B: Chemical, 178:485–493, 2013.\newline
[2]	Edward Song and Jin-Woo Choi. Conducting polyaniline nanowire and its applications in chemiresistive sensing. Nanomaterials, 3(3):498–523, 2013.\newline
[3]	Nirav Joshi, Maria Luisa Braunger, Flavio Makoto Shimizu, Antonio Riul, and Osvaldo Novais Oliveira. Two-dimensional transition metal dichalcogenides for gas sensing applications. Nanosens. Environ. Food Agric, 1:131–155, 2020.\newline
[4]	Yen-yu Chen. 2D MATERIALS FOR GAS-SENSING APPLICATIONS. PhD thesis, Purdue University Graduate School, 2021.\newline
[5]	Sajad Pirsa. Chemiresistive gas sensors based on conducting polymers. In Handbook of Research on Nanoelectronic Sensor Modeling and Applications, pages 150–180. IGI Global, 2017.\newline
[6]	Ananya Dey. Semiconductor metal oxide gas sensors: A review. Materials Science and Engineering: B, 229:206–217, 2018.\newline
[7]	Norizan M Nurazzi, Norli Abdullah, Siti ZN Demon, Norhana A Halim, Ahmad FM Azmi, Victor F Knight, and Imran S Mohamad. The frontiers of functionalized graphene-based nanocomposites as chemical sensors. Nanotechnology Reviews, 10(1):330–369, 2021.\newline
[8]	Jamal Malallah Rzaij and Amina Mohsen Abass. Review on: Tio2 thin film as a metal oxide gas sensor. Journal of Chemical Reviews, 2(2):114–121, 2020.\newline
[9]	Punit Tyagi, Anjali Sharma, Monika Tomar, and Vinay Gupta.   Metal oxide catalyst assisted sno2 thin film based so2 gas sensor. Sensors and Actuators B: Chemical, 224:282–289, 2016.\newline
[10]	Ming Yang, Yanyan Wang, Lei Dong, Zhiyong Xu, Yanhua Liu, Nantao Hu, Eric SiuWai Kong, Jiang Zhao, and Changsi Peng. Gas sensors based on chemically reduced holey graphene oxide thin films. Nanoscale research letters, 14(1):1–8, 2019.\newline
[11]	Irina V Zaporotskova, Natalia P Boroznina, Yuri N Parkhomenko, and Lev V Kozhitov. Carbon nanotubes: Sensor properties. a review. Modern Electronic Materials, 2(4):95– 105, 2016.\newline
[12]	Praveen Kumar Sahu, Rajiv K Pandey, R Dwivedi, VN Mishra, and R Prakash. Polymer/graphene oxide nanocomposite thin film for no 2 sensor: An in situ investigation of electronic, morphological, structural, and spectroscopic properties. Scientific reports, 10(1):1–13, 2020.\newline
[13]	Natalia Alzate-Carvajal and Adina Luican-Mayer. Functionalized graphene surfaces for selective gas sensing. ACS omega, 5(34):21320–21329, 2020.\newline
[14]	Kyriaki Kalaitzidou, Hiroyuki Fukushima, and Lawrence T Drzal. Multifunctional polypropylene composites produced by incorporation of exfoliated graphite nanoplatelets. Carbon, 45(7):1446–1452, 2007.\newline
[15]	Sumin Kim, Inhwan Do, and Lawrence T Drzal. Multifunctional xgnp/lldpe nanocomposites prepared by solution compounding using various screw rotating systems. Macromolecular Materials and Engineering, 294(3):196–205, 2009.\newline
[16]	Albert F. Rigosi, et al. Examining epitaxial graphene surface conductivity and quantum Hall device stability with Parylene passivation. Microelectronic engineering, 194:51-55, 2018.\newline
[17]	Albert F Rigosi, et al. Atypical quantized resistances in millimeter-scale epitaxial graphene pn junctions. Carbon, 154:230–237, 2019.\newline
[18]	Heather M Hill, et al. Probing the dielectric response of the interfacial buffer layer in epitaxial graphene via optical spectroscopy. Physical Review B, 96(19):195437, 2017.\newline
[19]	Yu-Sung Chang, Feng-Kuan Chen, Du-Cheng Tsai, Bing-Hau Kuo, and Fuh-Sheng Shieu. N-doped reduced graphene oxide for room-temperature no gas sensors. Scientific Reports, 11(1):1–12, 2021.\newline
[20]	Olasunbo Z Farinre, Hawazin Alghamdi, Mathew L Kelley, Adam J Biacchi, Albert V Davydov, Christina A Hacker, Albert F Rigosi, and Prabhakar Misra. A comprehensive study on the spectroscopic characterization and molecular dynamics simulation of pristine and functionalized graphene nanoplatelets for gas sensing applications. arXiv preprint arXiv:2201.04683, 2022.\newline
[21]	Haifeng Fei, Gang Wu, Wei-Ying Cheng, Wenjie Yan, Hongjun Xu, Duan Zhang, Yanfeng Zhao, Yanhui Lv, Yanhui Chen, Lei Zhang, et al. Enhanced no2 sensing at room temperature with graphene via monodisperse polystyrene bead decoration. ACS omega, 4(2):3812–3819, 2019.\newline
[22]	Dmytro Kostiuk, Stefan Luby, Peter Siffalovic, Monika Benkovicova, Jan Ivanco, Matej Jergel, and Eva Majkova. Graphene langmuir-schaefer films decorated by pd nanoparticles for no 2 and h 2 gas sensors. Measurement Science Review, 19(2):64–69, 2019.\newline
[23]	Tao Wang, Da Huang, Zhi Yang, Shusheng Xu, Guili He, Xiaolin Li, Nantao Hu, Guilin Yin, Dannong He, and Liying Zhang. A review on graphene-based gas/vapor sensors with unique properties and potential applications. Nano-Micro Letters, 8(2):95–119, 2016.\newline
[24]	A Jim´enez-Su´arez and SG Prolongo.  Graphene nanoplatelets, 2020.\newline
[25]	COMSOL  Multiphysics.    Introduction  to  comsol  multiphysicsQR .    COMSOL  Multiphysics, Burlington, MA, accessed Feb, 9:2018, 1998.\newline
[26]	Wolfram Research. ElementData. https://reference.wolfram.com/language/ref/ ElementData.html, 2014.\newline
[27]	CB     Honsberg     and     SG     Bowden.	Photovoltaics education website, https://www.pveducation.org. PV Education, 2019.\newline
[28]	Michelle M Gauthier. Engineering materials handbook. ASM International, November, 1995.\newline
[29]	Richard S Muller, Theodore I Kamins, Mansun Chan, and Ping K Ko. Device electronics for integrated circuits. 1986.\newline
[30]	Frederick Seitz and RP Johnson. Modern theory of solids. ii. Journal of Applied Physics, 8(3):186–199, 1937.\newline
[31]	IN Shklyarevskii and PL Pakhomov. Separation of contributions from free and coupled electrons into real and imaginary parts of a dielectric-constant of gold. Optika i Spektroskopiya, 34(1):163–166, 1973.\newline
[32]	Andi Wang and DDL Chung. Dielectric and electrical conduction behavior of carbon paste electrochemical electrodes, with decoupling of carbon, electrolyte and interface contributions. Carbon, 72:135–151, 2014.\newline
[33]	Benyamin Davaji, Hak Dong Cho, Mohamadali Malakoutian, Jong-Kwon Lee, Gennady Panin, Tae Won Kang, and Chung Hoon Lee. A patterned single layer graphene resistance temperature sensor. Scientific reports, 7(1):1–10, 2017.\newline
[34]	Xiaohui Tang, Marc Debliquy, Driss Lahem, Yiyi Yan, and Jean-Pierre Raskin. A review on functionalized graphene sensors for detection of ammonia. Sensors, 21(4):1443, 2021.\newline
[35]	Listya Utari, Ni Luh Wulan Septiani, Levy Olivia Nur, Hutomo Suryo Wasisto, Brian Yuliarto, et al. Wearable carbon monoxide sensors based on hybrid graphene/zno nanocomposites. IEEE Access, 8:49169–49179, 2020.\newline
[36] Jiuning Hu, et al. Quantum transport in graphene p− n junctions with moiré superlattice modulation. Physical Review B, 98(4):045412, 2018.\newline

\pagebreak

\bibliographystyle{unsrt}
\bibliography{REFERENCES.bib}

\begin{thebibliography}{10}

\bibitem{wu2013enhanced}
Zuquan Wu, Xiangdong Chen, Shibu Zhu, Zuowan Zhou, Yao Yao, Wei Quan, and Bin
  Liu.
\newblock Enhanced sensitivity of ammonia sensor using graphene/polyaniline
  nanocomposite.
\newblock {\em Sensors and Actuators B: Chemical}, 178:485--493, 2013.

\bibitem{song2013conducting}
Edward Song and Jin-Woo Choi.
\newblock Conducting polyaniline nanowire and its applications in
  chemiresistive sensing.
\newblock {\em Nanomaterials}, 3(3):498--523, 2013.

\bibitem{joshi2020two}
Nirav Joshi, Maria~Luisa Braunger, Flavio~Makoto Shimizu, Antonio Riul, and
  Osvaldo~Novais Oliveira.
\newblock Two-dimensional transition metal dichalcogenides for gas sensing
  applications.
\newblock {\em Nanosens. Environ. Food Agric}, 1:131--155, 2020.

\bibitem{chen20212d}
Yen-yu Chen.
\newblock {\em 2D MATERIALS FOR GAS-SENSING APPLICATIONS}.
\newblock PhD thesis, Purdue University Graduate School, 2021.

\bibitem{pirsa2017chemiresistive}
Sajad Pirsa.
\newblock Chemiresistive gas sensors based on conducting polymers.
\newblock In {\em Handbook of Research on Nanoelectronic Sensor Modeling and
  Applications}, pages 150--180. IGI Global, 2017.

\bibitem{dey2018semiconductor}
Ananya Dey.
\newblock Semiconductor metal oxide gas sensors: A review.
\newblock {\em Materials Science and Engineering: B}, 229:206--217, 2018.

\bibitem{nurazzi2021frontiers}
Norizan~M Nurazzi, Norli Abdullah, Siti~ZN Demon, Norhana~A Halim, Ahmad~FM
  Azmi, Victor~F Knight, and Imran~S Mohamad.
\newblock The frontiers of functionalized graphene-based nanocomposites as
  chemical sensors.
\newblock {\em Nanotechnology Reviews}, 10(1):330--369, 2021.

\bibitem{rzaij2020review}
Jamal~Malallah Rzaij and Amina~Mohsen Abass.
\newblock Review on: Tio2 thin film as a metal oxide gas sensor.
\newblock {\em Journal of Chemical Reviews}, 2(2):114--121, 2020.

\bibitem{tyagi2016metal}
Punit Tyagi, Anjali Sharma, Monika Tomar, and Vinay Gupta.
\newblock Metal oxide catalyst assisted sno2 thin film based so2 gas sensor.
\newblock {\em Sensors and Actuators B: Chemical}, 224:282--289, 2016.

\bibitem{yang2019gas}
Ming Yang, Yanyan Wang, Lei Dong, Zhiyong Xu, Yanhua Liu, Nantao Hu, Eric
  Siu-Wai Kong, Jiang Zhao, and Changsi Peng.
\newblock Gas sensors based on chemically reduced holey graphene oxide thin
  films.
\newblock {\em Nanoscale research letters}, 14(1):1--8, 2019.

\bibitem{zaporotskova2016carbon}
Irina~V Zaporotskova, Natalia~P Boroznina, Yuri~N Parkhomenko, and Lev~V
  Kozhitov.
\newblock Carbon nanotubes: Sensor properties. a review.
\newblock {\em Modern Electronic Materials}, 2(4):95--105, 2016.

\bibitem{sahu2020polymer}
Praveen~Kumar Sahu, Rajiv~K Pandey, R~Dwivedi, VN~Mishra, and R~Prakash.
\newblock Polymer/graphene oxide nanocomposite thin film for no 2 sensor: An in
  situ investigation of electronic, morphological, structural, and
  spectroscopic properties.
\newblock {\em Scientific reports}, 10(1):1--13, 2020.

\bibitem{alzate2020functionalized}
Natalia Alzate-Carvajal and Adina Luican-Mayer.
\newblock Functionalized graphene surfaces for selective gas sensing.
\newblock {\em ACS omega}, 5(34):21320--21329, 2020.

\bibitem{kalaitzidou2007multifunctional}
Kyriaki Kalaitzidou, Hiroyuki Fukushima, and Lawrence~T Drzal.
\newblock Multifunctional polypropylene composites produced by incorporation of
  exfoliated graphite nanoplatelets.
\newblock {\em Carbon}, 45(7):1446--1452, 2007.

\bibitem{kim2009multifunctional}
Sumin Kim, Inhwan Do, and Lawrence~T Drzal.
\newblock Multifunctional xgnp/lldpe nanocomposites prepared by solution
  compounding using various screw rotating systems.
\newblock {\em Macromolecular Materials and Engineering}, 294(3):196--205,
  2009.

\bibitem{rigosi2019gateless}
Albert~F Rigosi, Mattias Kruskopf, Heather~M Hill, Hanbyul Jin, Bi-Yi Wu,
  Philip~E Johnson, Siyuan Zhang, Michael Berilla, Angela R~Hight Walker,
  Christina~A Hacker, et~al.
\newblock Gateless and reversible carrier density tunability in epitaxial
  graphene devices functionalized with chromium tricarbonyl.
\newblock {\em Carbon}, 142:468--474, 2019.

\bibitem{rigosi2019atypical}
Albert~F Rigosi, Dinesh Patel, Martina Marzano, Mattias Kruskopf, Heather~M
  Hill, Hanbyul Jin, Jiuning Hu, Angela R~Hight Walker, Massimo Ortolano, Luca
  Callegaro, et~al.
\newblock Atypical quantized resistances in millimeter-scale epitaxial graphene
  pn junctions.
\newblock {\em Carbon}, 154:230--237, 2019.

\bibitem{rigosi2019quantum}
Albert~F Rigosi and Randolph~E Elmquist.
\newblock The quantum hall effect in the era of the new si.
\newblock {\em Semiconductor science and technology}, 34(9):093004, 2019.

\bibitem{chang2021n}
Yu-Sung Chang, Feng-Kuan Chen, Du-Cheng Tsai, Bing-Hau Kuo, and Fuh-Sheng
  Shieu.
\newblock N-doped reduced graphene oxide for room-temperature no gas sensors.
\newblock {\em Scientific Reports}, 11(1):1--12, 2021.

\bibitem{farinre2022comprehensive}
Olasunbo~Z Farinre, Hawazin Alghamdi, Mathew~L Kelley, Adam~J Biacchi, Albert~V
  Davydov, Christina~A Hacker, Albert~F Rigosi, and Prabhakar Misra.
\newblock A comprehensive study on the spectroscopic characterization and
  molecular dynamics simulation of pristine and functionalized graphene
  nanoplatelets for gas sensing applications.
\newblock {\em arXiv preprint arXiv:2201.04683}, 2022.

\bibitem{fei2019enhanced}
Haifeng Fei, Gang Wu, Wei-Ying Cheng, Wenjie Yan, Hongjun Xu, Duan Zhang,
  Yanfeng Zhao, Yanhui Lv, Yanhui Chen, Lei Zhang, et~al.
\newblock Enhanced no2 sensing at room temperature with graphene via
  monodisperse polystyrene bead decoration.
\newblock {\em ACS omega}, 4(2):3812--3819, 2019.

\bibitem{kostiuk2019graphene}
Dmytro Kostiuk, Stefan Luby, Peter Siffalovic, Monika Benkovicova, Jan Ivanco,
  Matej Jergel, and Eva Majkova.
\newblock Graphene langmuir-schaefer films decorated by pd nanoparticles for no
  2 and h 2 gas sensors.
\newblock {\em Measurement Science Review}, 19(2):64--69, 2019.

\bibitem{wang2016review}
Tao Wang, Da~Huang, Zhi Yang, Shusheng Xu, Guili He, Xiaolin Li, Nantao Hu,
  Guilin Yin, Dannong He, and Liying Zhang.
\newblock A review on graphene-based gas/vapor sensors with unique properties
  and potential applications.
\newblock {\em Nano-Micro Letters}, 8(2):95--119, 2016.

\bibitem{jimenez2020graphene}
A~Jim{\'e}nez-Su{\'a}rez and SG~Prolongo.
\newblock Graphene nanoplatelets, 2020.

\bibitem{multiphysics1998introduction}
COMSOL Multiphysics.
\newblock Introduction to comsol multiphysics{\textregistered}.
\newblock {\em COMSOL Multiphysics, Burlington, MA, accessed Feb}, 9:2018,
  1998.

\bibitem{reference.wolfram_2021_elementdata}
Wolfram Research.
\newblock {ElementData}.
\newblock \url{https://reference.wolfram.com/language/ref/ElementData.html},
  2014.

\bibitem{honsberg2019photovoltaics}
CB~Honsberg and SG~Bowden.
\newblock Photovoltaics education website, https://www.pveducation.org.
\newblock {\em PV Education}, 2019.

\bibitem{gauthier1995engineering}
Michelle~M Gauthier.
\newblock Engineering materials handbook.
\newblock {\em ASM International, November}, 1995.

\bibitem{muller1986device}
Richard~S Muller, Theodore~I Kamins, Mansun Chan, and Ping~K Ko.
\newblock Device electronics for integrated circuits.
\newblock 1986.

\bibitem{seitz1937modern}
Frederick Seitz and RP~Johnson.
\newblock Modern theory of solids. ii.
\newblock {\em Journal of Applied Physics}, 8(3):186--199, 1937.

\bibitem{shklyarevskii1973separation}
IN~Shklyarevskii and PL~Pakhomov.
\newblock Separation of contributions from free and coupled electrons into real
  and imaginary parts of a dielectric-constant of gold.
\newblock {\em Optika i Spektroskopiya}, 34(1):163--166, 1973.

\bibitem{wang2014dielectric}
Andi Wang and DDL Chung.
\newblock Dielectric and electrical conduction behavior of carbon paste
  electrochemical electrodes, with decoupling of carbon, electrolyte and
  interface contributions.
\newblock {\em Carbon}, 72:135--151, 2014.

\bibitem{davaji2017patterned}
Benyamin Davaji, Hak~Dong Cho, Mohamadali Malakoutian, Jong-Kwon Lee, Gennady
  Panin, Tae~Won Kang, and Chung~Hoon Lee.
\newblock A patterned single layer graphene resistance temperature sensor.
\newblock {\em Scientific reports}, 7(1):1--10, 2017.

\bibitem{tang2021review}
Xiaohui Tang, Marc Debliquy, Driss Lahem, Yiyi Yan, and Jean-Pierre Raskin.
\newblock A review on functionalized graphene sensors for detection of ammonia.
\newblock {\em Sensors}, 21(4):1443, 2021.

\bibitem{utari2020wearable}
Listya Utari, Ni~Luh~Wulan Septiani, Levy~Olivia Nur, Hutomo~Suryo Wasisto,
  Brian Yuliarto, et~al.
\newblock Wearable carbon monoxide sensors based on hybrid graphene/zno
  nanocomposites.
\newblock {\em IEEE Access}, 8:49169--49179, 2020.

\end{thebibliography}

\end{document}